\documentclass[12pt]{article}
\usepackage{epsfig}

\usepackage{a4}
\usepackage[usenames]{color}

\usepackage{amsmath}

\usepackage{amssymb}
\usepackage{graphicx}
\usepackage{amsmath}
\usepackage{amsfonts} 
\usepackage{xcolor}

\newcommand{\la}{\lambda}

\newcommand{\ka}{\kappa}
\newcommand{\f}{\phi}
\newcommand{\F}{\Phi}
\newcommand{\vf}{\varphi}
\newcommand{\ta}{\theta}
\newcommand{\Ta}{\Theta}
\newcommand{\al}{\alpha}
\newcommand{\bt}{\beta}

\newcommand{\ee}{\end{equation}}
\newcommand{\eea}{\end{eqnarray}}
\newcommand{\be}{\begin{equation}}
\newcommand{\bea}{\begin{eqnarray}}

\newcommand{\pa}{\partial}
\newcommand{\Om}{\Omega}
\newcommand{\vep}{\varepsilon}
\newcommand{\vr}{\varrho}

\newcommand{\om}{\omega}
\def\theequation{\arabic{equation}}

\newcommand{\re}[1]{(\ref{#1})}
\newcommand{\R}{{\rm I \hspace{-0.52ex} R}}

\numberwithin{equation}{section}
\begin{document}

\title{
\bf 
The effect of Skyrme--Chern-Simons dynamics on gauged Skyrmions in $2+1$ dimensions
}
\author{ {\large Francisco Navarro-L\'erida}$^{\star}$, 
{\large Eugen Radu}$^{\diamond}$ 
and 
{\large D. H. Tchrakian}$^{\dagger **}$
\\ 
\\
$^{\star}${\small  
Departamento de F\'isica Te\'orica and IPARCOS, Ciencias F\'isicas,
}\\
{\small Universidad Complutense de Madrid, E-28040 Madrid, Spain} 
\\  
$^{\diamond}${\small
 Departamento de Matem\'atica da Universidade de Aveiro and }
\\
{\small 
Center for Research and Development in Mathematics and Applications,}
\\
{\small 
 Campus de Santiago, 3810-183 Aveiro, Portugal}
\\ 
$^{\dagger}${\small 
School of Theoretical Physics, Dublin Institute for Advanced Studies,}
\\
{\small  Burlington Road, Dublin 4, Ireland}
\\   
$^{**}${\small Department of Computer Science, National University of Ireland Maynooth, Maynooth, Ireland }
}

\date{}

\maketitle


\bigskip

\begin{abstract}
We study the Skyrmion of the $SO(2)$ gauged $O(3)$ sigma model in $2+1$ dimensions in the presence of
a Skyrme--Chern-Simons (SCS) term, and compare its properties with the corresponding properties of the Skyrmion
in the presence of the usual Chern-Simons (CS) term. We find that these properties are qualitatively largely
similar in both cases, meaning that the SCS density can be employed as an alternative to the CS term also in
higher dimensions, most importantly in even dimensions where no CS term is defined, $e.g.,$ in
$3+1$ dimensions.
The SCS density employed here is defined in terms of the pair of $SO(2)\times SO(2)$ gauge fields and
an auxiliary $O(5)$ Skyrme scalar, which is contracted to an effective $O(3)$ Skyrme scalar.
Technically, this study maps the methods to be applied in higher dimensional examples.


\end{abstract}
\newpage

\tableofcontents

\section{Introduction and motivation}
The aim of this report is to test whether the 
effects of Chern-Simons (CS) dynamics on the solitons of gauged Skyrmions persist also when the
CS term in the Lagragian is replaced by a Skyrme--Chern-Simons ~\cite{Tchrakian:2015pka,Tchrakian:2021xzy}
(SCS) term.

Let us start with recalling the effects of CS dynamics on the solitons of gauged Skyrme 
models, against which it is our aim here to compare the effects of SCS dynamics.
The effects of CS dynamics on gauged Skyrmions in $2+1$ and $4+1$ dimensions were discovered in
Refs.~\cite{Navarro-Lerida:2016omj,Navarro-Lerida:2018giv,Navarro-Lerida:2018siw} and 
\cite{Navarro-Lerida:2020jft,Navarro-Lerida:2020hph}, respectively.
These were found to be
quite remarkable. Specifically, it turned out that the dependence of the static energy $E$ on
the electric charge $Q_e$ and the angular momentum $J$ was {\it non-standard}, in that the
slopes of $(E,Q_e)$ and $(E,J)$-curves have both {\it positive and negative} signs, in contrast with
the {\it standard} case of exclusively monotonically increasing {\it positive} slopes.
 In $2+1$ dimensions, such results pertaining to the slope of $(E,Q_e)$
were also found in the Maxwell--Chern-Simons system, as shown in Ref.~\cite{Samoilenka:2016wys}.

In addition to these effects, it was observed that the effective ``baryon number'' could depart from
the {\it winding number}~\footnote{We have verified since, that these effects persist also in the absence
of the Maxwell term, in a model with Chern-Simons dynamics only.
}. More precisely, it was
found that the topological charge, the ``baryon number'', was altered due to the effect of
the gauge field in the presence of the CS term, on the (ungauged) Skyrmion.
It may be reasonable to conclude that these effects will persist for Abelian gauged Skyrme models
in all  odd dimensional spacetimes, where a CS density is defined.


Then, there arises the natural question as to whether such effects may be present as well
for gauged Skyrmions in even dimensional spacetimes where no CS density is defined,
for a Lagrangian featuring a SCS term instead. We believe this is true since in the $2+1$ dimensional $SO(2)$
gauged $O(3)$ Skyrme model featuring a SCS term studied here, the qualitative effects observed in the same
model subject instead to CS dynamics are largely replicated. Specifically, both {\it positive and negative}
slopes of $(E,Q_e)$ and $(E,J)$-curves are observed. The change of ``baryon number'' resulting from CS dynamics
is however not observed here for SCS dynamics. This disparity between CS and SCS however,
is peculiar to $2+1$ dimensions only and hence does not present an obstacle to asserting that CS and SCS
lead to qualitatively similar effects in general.



\medskip

Let us start with the prescription for $SO(N)$ gauging of an $O(D+1)$ Skyrme scalar $\f^a$ on $\R^D$.
Prior to gauging, the Skyrmion described by $\f^a\ ,\ a=1,2,\dots,D+1$ subject to
$|\f^a|^2=1$ on $\R^D$, is stabilised by the topological charge (``baryon number''),
namely the volume integral of the winding number density denoted by $\vr^{(D)}_0$. The density $\vr^{(D)}_0$
presents a lower bound on the energy density.
After gauging~\footnote{Note that here the choice of gauge group is
$SO(N)\ ,\ 2\le N\le D$, unlike in the corresponding gauged Higgs system where only one gauge group is
consistent with the finiteness of energy.} with $SO(N)\ ,\ 2\le N\le D$, the gauge covariant derivatives read
\bea
D_i\f^\al&=&\pa_i\f^\al+A_i^{\al\bt}\f^\bt\ ,\quad\al=1,2,\dots,N\ ;\quad A_i^{\al\bt}\in so(N)\label{coval}\\
D_i\f^A&=&\pa_i\f^A\ ,\quad A=N+1, N+2,\dots,D+1 \ , \label{covA}
\eea
the index $i=1,2,\dots,D$ labelling the coordinate on $\R^D$.

Starting from the standard definition of the winding number density $\vr_0^{(D)}$ of the $O(D+1)$ Skyrme scalar
in $D$ dimensions, one can define a gauge invariant density $\vr_G^{(D)}$ by replacing all partial derivatives of the Skyrme
scalar $\f^a$ in $\vr_0^{(D)}$ by the corresponding covariant derivatives~\re{coval}-\re{covA},
which however is not a total divergence like $\vr_0^{(D)}$. The required energy density lower bound must be
both gauge-invariant and total-divergence. Such a density can be constructed by calculating the difference
$(\vr_G^{(D)}-\vr_0^{(D)})$
\be
\label{vrgminusvr0}
\vr_G^{(D)}-\vr_0^{(D)}=\pa_i \Om_i^{(D)}[A,\pa\f]-W[F,D\f] \,,
\ee
where $W[F,D\f]$ is constructed to be gauge invariant by isolating all total-divergence quantities
$\pa_i\Om_i^{(D)}[A,\pa\f]$. The relation \re{vrgminusvr0} is just a prescription, which must be calculated for
each dimension individually~\footnote{To date, this prescription is implemented explicitly in $D=2,3$ and $4$
with gauge group $SO(D)$ and all its contractions.
}.
The required gauge-invariant and total-divergence density then is defined by the following two rearrangements
of \re{vrgminusvr0}
\bea
\vr^{(D)}&=&\vr_G^{(D)}+W[F,D\f]\label{vr1}\\
&=&\vr_0^{(D)}+\pa_i\Om_i^{(D)}[A,\pa\f]\,,\label{vr2}
\eea
of which the version \re{vr1} being manifestly gauge invariant and the version \re{vr2} being manifestly
total-divergence, since $\vr^{(D)}_0$ is (essentially) total-divergence. 
As such, the density \re{vr1}-\re{vr2} is a gauged Skyrme analogue of the usual Chern-Pontryagin (CP)
 density, albeit for a given orthogonal gauge group, and for which its definition is contingent on an explicit 
construction in each dimension.

In the construction of gauged Skyrmions, the version \re{vr1} of $\vr^{(D)}$ is exploited to devise Bogomol'nyi
type energy lower bounds while the version \re{vr2} can serve in the evaluation of this bound by calculating the
relevant ``surface integral''. In the absence of the CS term, the latter will be the integer ``baryon number'',
while in the presence of the CS term it can depart from this. For the case at hand, where the Skyrmion on $\R^2$
is concerned, \re{vr2} is given explicitly and a discussion why in this model ``baryon number'' change is not
demonstrated.

A different application of $\vr^{(D)}$ arises from the fact that it is an analogue of the CP density
which likewise is gauge-invariant and total-divergence. Just as the total-divergence expression of the CP density
enables the definition of the CS density by a one-step descent, applying this one-step descent to \re{vr2} results
in the definition of the SCS density in $D-1$ ``spacetime'' dimensions.
Expressing $\vr_0^{(D)}=\pa_i\om_i$ in \re{vr2}, this descent is achieved by setting $i=D$, with the SCS density defined
formally by
\bea
\Om_{\rm SCS}=\om_{i=D}+\Om_{i=D}^{(D)}
\eea
which is presented in Appendix {\bf A}.

The salient difference between the CS and the SCS densities is that the former are defined only in odd
dimensions being descended by one step from the CP densities defined only in even dimensions, while the latter
result from a one-step descent of the densities \re{vr2}, defined in all dimensions, so that the SCS
densities are also defined in all dimensions.

There is yet another important difference between CS and SCS densities. While the CS density is defined
exclusively in terms of the gauge field $A_\mu$, the SCS density is described by $A_\mu$, the auxiliary gauge
field $B_\mu$, and the auxiliary $O(5)$ scalar $\ta^{\bar a}$. This
property, namely that the SCS density is defined in terms of both the gauge field(s) {\bf and} the Skyrme scalar, is shared
with the Callan-Witten~\cite{Callan:1983nx} anomaly.


In the present work in $2+1$ dimensions, we consider  only the SCS density {\it in the absence} of the usual CS density,
with the purpose of finding out to what extent the dynamical effects of the CS term studied in previous
investigations~\cite{Navarro-Lerida:2016omj,Navarro-Lerida:2018giv,Navarro-Lerida:2018siw,Navarro-Lerida:2020jft,Navarro-Lerida:2020hph}
are repeated qualitatively when the SCS density is employed. As such the present study serves the purpose of a prototype. At this
preliminary stage we have chosen to consider a system in $2+1$ dimensions for a technical reason,
namely that the considerable task of
numerical constructions can be carried out by solving ODE's, rather than PDE's.


This paper is organized as follows.
In Section {\bf 2} the model is presented, which includes the detailed structure of the SCS density, as well as the
(sigma model) constraint compliant parametrisation of the system, which is instrumental in the definition 
of global charges.  
Symmetry imposition is presented in Section {\bf 3}. 
There also the gauge field equations are stated
and the electric charges and angular momenta are calculated algebraically.  
The results of numerical construction of the solutions for several
 contractions of the general model are presented
in detail in Section {\bf 4}. 
A summary of results is given in Section {\bf 5} and natural extensions of
SCS densities to $3+1$ dimensions are projected for future consideration. A brief presentation of the SCS
term employed, is given in Appendix {\bf A}.

\medskip
 {\bf Conventions.}
%
The background of the theory is a three dimensional 
Minkowski spacetime (with the signature $(+--)$). 
Its spatial $\mathbb{R}^2$ part is written
first  in terms of 
Cartesian coordinate, 
$ds_2^2=dx_1^2+dx_2^2$,
with the equivalent form  
$ds_2^2=d r^2+r^2 d\varphi^2 $
obtained via the usual transformation
$x_1=r \cos \varphi$,
$x_2=r \sin \varphi$, 
where
$0\leq r<\infty$, 
and 
$0\leq \varphi <2\pi$. 

Also, throughout the paper,
early Latin letters, $a,b,\dots $ label the internal indices of the Skyrme fields
while
 Greek alphabet letters $\alpha,\beta,\dots$ 
label spacetime coordinates, 
  running from $0$ to $2$ (with $x^0=t$).
The Einstein's summation convention is used, 
although no distinction is made between covariant and contravariant \textit{internal} indices.

\section{The general model}
The basic system under study is the $SO(2)$ gauged $O(3)$ sigma model in $2+1$ dimensions, whose solitons
(Skyrmions) are influenced by the SCS dynamics.

The Lagrangian of the system we study here is
\be
\label{L}
{\cal L}={\cal L}_{(A_\mu,\f^a)}+{\cal L}_{(A_\mu,B_\mu,\ta^{\bar a})} \,,
\ee
where
\be
\label{Af}
{\cal L}_{(A_\mu,\f^a)}=-\frac14\, F_{\mu\nu}^2
+\frac12\,\eta^2\,|D_\mu\f^a|^2-\eta^4\, V[\f^a] \,,
\ee
is a simplified version of the $SO(2)$ gauged $O(3)$ sigma model studied in \cite{Navarro-Lerida:2018siw}
described by the kinetic term $F_{\mu\nu}^2=(\pa_\mu A_\nu-\pa_\nu A_\mu)^2$ of the ``Maxwell'' field $A_\mu$
and the quadratic kinetic term of $|D_\mu\f^a|^2$ of the $O(3)$ Skyrme scalar
$\f^{a}=(\f^\al,\f^3)\ ,\ \al=1,2$ subject to $|\f^a|^2=1$.
The constant $\eta$ in \re{Af} has dimension $L^{-1}$, keeping track of dimensions.

The gauging prescription given by the definition of the covariant derivative~\cite{Navarro-Lerida:2018siw} is
\bea
D_\mu\f^\al&=&\pa_\mu\f^\al+A_\mu\,(\vep\f)^\al \,, \label{Dfal}\\
D_\mu\f^3&=&\pa_\mu\f^3\,.\label{Df3-0}
\eea
In \re{Dfal}, the notation $(\vep\f)^\al=\vep^{\al\bt}\f^\bt$ is used. This notation is used also 
for the components $\ta^\al$ and $\ta^A$ below.

The system \re{Af} supports the $SO(2)$ gauged
planar Skyrmion~\cite{Schroers:1995he}, the Abelian field there being the ``physical'' Maxwell field. Previously
in Ref.~\cite{Navarro-Lerida:2016omj}, the CS term was added to
an extended version of \re{Af} which included the quartic kinetic Skyrme term
and its dynamical infuence on the Skyrmion was studied quantatively. Here, we have eschewed use of the
quartic kinetic Skyrme term since its presence is not required by Derrick scaling~\footnote{The 
quartic kinetic Skyrme term was included in Ref.~\cite{Navarro-Lerida:2016omj} because there, we were interested in the gauge
decoupling limit, which we certainly are not interested in here.}.

In the present work we intend to replace the CS term by a SCS density, to study the influence
of the latter on the gauged planar Skyrmion. To this end, we have introduced 
the second term in the full Lagrangian \re{L}
\be
\label{Lta}
{\cal L}_{(A_\mu,B_\mu,\ta^{\bar a})}=
\ka\,\Om_{\rm SCS}-\frac14\,G_{\mu\nu}^2+\frac12\,\eta^2\,|D_\mu\ta^{\bar a}|^2
-\eta^4\,V[\ta^{\bar a}] \,,
\ee
in which $G_{\mu\nu}^2=(\pa_\mu B_\nu-\pa_\nu B_\mu)^2$ is the kinetic term of the ``auxiliary''
Abelian field $B_\mu$, and $|D_\mu\ta^{\bar a}|^2$ is the quadratic kinetic term of the $O(5)$
"auxiliary" Skyrme scalar $\ta^{\bar a}=(\ta^\al,\ta^A,\ta^5)$
whose gauging prescription is given by the covariant derivatives
\bea
D_\mu\ta^\al&=&\pa_\mu\ta^\al+A_\mu\,(\vep\ta)^\al \,, \ \ \ \alpha=1,2 \,,
\label{Dtaal}
\\
D_\mu\ta^A&=&\pa_\mu\ta^A+B_\mu\,(\vep\ta)^A \,, \ \ \ A=3,4 \,,
\label{DtaA0}
\\
D_\mu\ta^5&=&\pa_\mu\ta^5\,.\label{Dta5}
\eea
$V[\f^a]$ and $V[\ta^{\bar a}]$ in \re{Af} and \re{Lta}, respectively, are potential terms which are in
practice taken to be the ``pion mass'' potentials
\be
V[\f^a]=\mu (1-\f^3)\quad{\rm and}\quad V[\ta^{\bar a}]=\mu (1-\ta^5)\, ,\label{pots}
\ee
$\mu$ being a constant.

It may be in order here to comment on the choice of the ``pion mass'' potential for the Skyrme scalar $\f^a$, in \re{pots}. Ideally, it would be desirable to have a potential like
\be
V\simeq|\f^\al|^2(\f^3-\upsilon)^2\ ,\quad\al=1,2 \ , \label{AppC}
\ee
with $\upsilon \in [0,1]$, which  in the ``usual'' CS model enabled the construction of solutions describing change of
``baryon number'', as shown in Appendix {\bf C} of Ref.~\cite{Navarro-Lerida:2018siw}. Unfortunately,
we have been unable to construct such a potential when replacing the CS term by the SCS term employed here.
  In Appendix {\bf B} of Ref.~\cite{Navarro-Lerida:2018siw}, the stability requirement for Maxwell gauged $O(3)$
Skyrmions on $\R^2$ led to the potential
\be
\label{Schpot}
V[\f^a]\simeq(1-\f^3)^2\,,
\ee
which is the potential that allows for the self-dual solutions of Ref.~\cite{Schroers:1995he}. But since the function
$(1-\f^3)$ is everywhere larger than the function $(1-\f^3)^2/2$, replacing the potential \re{Schpot} with the
``pion mass'' potential in \re{pots} does not violate the criterion of topological stability, legitimising our use
 of the potential \re{pots} here. By contrast, we cannot replace the potential \re{Schpot} with \re{AppC},
since the latter is {\bf not} everywhere larger that \re{Schpot}.

Having opted to adopt the first member of \re{pots} as the potential in \re{Af}, namely that
\be
\label{pot}
V[\f^a]=\mu (1-\f^3)\,,
\ee
it follows that the lower bound density \re{vr2}, which in two dimensions takes the explicit form
\bea
\vr^{(2)}&=&\vr_0^{(2)}+2\vep_{ij}\,\pa_i[A_j(\f^3-1)]\,,\label{vr2x}
\eea
and since $\displaystyle \lim_{r\to\infty}\f^3=1$, gives rise to an integer ``baryon number'', namely the integral of $\vr_0^{(2)}$, since the surface (line) integral of the second term in  \re{vr2x} vanishes.


The most important term in \re{Lta} is the SCS density, $\Om_{\rm SCS}$, which here replaces
the CS term employed previously in Refs.~\cite{Navarro-Lerida:2016omj,Navarro-Lerida:2018giv,Navarro-Lerida:2018siw}.
In addition to the ``physical'' Maxwell field $A_\mu$, the SCS
density depends also on the auxiliary Abelian field $B_\mu$ and the $O(5)$ auxiliary Skyrme scalar 
$\ta^{\bar a}$, so the terms $G_{\mu\nu}^2$ and $|D_\mu\ta^{\bar a}|^2$ have been included in \re{Lta} to confer proper
dynamics to these auxiliary fields~\footnote{We have recently verified numerically that the dynamical
effects observed with the CS term persist also with the SCS term, also in the absence of the $G_{\mu\nu}^2$
in the Lagrangian \re{Lta}.}.

The $SO(2)\times SO(2)$ SCS density in $2+1$ spacetime employed here is given in detail in Section {\bf 4.2}
of Ref.~\cite{Tchrakian:2021xzy} and in Appendix {\bf A} here. Denoting it as
\be
\label{SCS}
\Om_{\rm SCS}=\om+\Om\,,
\ee
it consists of two terms.

The first term, $\om$, is the Wess-Zumino term extracted from $\vr_0$, which 
can be displayed explicitly in {\it constraint compliant} parametrisation. But since in the static limit
$\om$ vanishes by symmetry, it will not play a role in the construction of these solutions.

The second term, $\Om$, is
\bea
\Om&=&3!\eta\,\vep^{\mu\nu\la}\ \ta^5\bigg\{\frac13\,(\ta^5)^2\nonumber
(A_\la G_{\mu\nu}+B_\la F_{\mu\nu})\nonumber\\
&&\qquad\qquad\quad-A_\mu B_\nu\ \pa_\la(|\ta^\al|^2-|\ta^A|^2)\nonumber\\
&&\qquad\qquad\quad
-2\left[A_\la\ (\vep\pa_\mu\ta)^A\,\pa_\nu\ta^A+B_\la\ (\vep\pa_\mu\ta)^\al\,\pa_\nu\ta^\al\right]\bigg\}\,.\label{SCSAB}
\eea

This completes the formal definition of the model, which in addition to two Abelian gauge 
fields $A_\mu$ and $B_\mu$, also features 
an $O(3)$ Skyrme scalar $\f^a$ and an $O(5)$ Skyrme scalar $\ta^{\bar a}$. Of these, $(A_\mu,\f^a)$ describe
the gauged Skyrmion (soliton) which has an energy lower bound, while $(A_\mu,B_\mu,\ta^{\bar a})$ describe
the SCS density $\Om_{\rm SCS}$, \re{SCS} which influences the Skyrmion dynamics in the way that the
CS density does.

The general model proposed here will be contracted systematically, but since the SCS density is of
intrinsic interest, we have relegated these contractions to Section {\bf 2.2}, after expressing the
model in constraint compliant parametrisation in Section {\bf 2.1}.

\subsection{Constraint compliant parametrisation}
The constraint compliant parametrisation of the
Skyrme scalar $\f^a=(\f^\al,\f^3)$ is expressed by
\be
\label{ccf}
\f^{a}=\left(\begin{array}{l}
\f^\al\\
\f^3
\end{array}\right)
=\left(\begin{array}{l}
\sin h(x_\mu)\,n^{\al}\\
\cos h(x_\mu)
  \end{array}\right)\equiv
\left(\begin{array}{l}
\F_1(x_\mu)\,n^{\al}\\
\F_2(x_\mu)\end{array}\right) \,,
\ee
and the $O(5)$ Skyrme scalar $\ta^{\bar a}=(\ta^\al,\ta^A,\ta^5)\ ,\ \al=1,2;\ A=3,4$ is parametrised by
\be
\label{cct}
\ta^{\bar a}=\left(\begin{array}{l}
\ta^{\al}\\
\ta^{A}\\
\ta^{5}
\end{array}\right)
=
\left(\begin{array}{l}
 \sin f(x_\mu)\sin g(x_\mu)\ n^{\al}\\
\sin f(x_\mu)\cos g(x_\mu)\ m^{A}\\
\cos f(x_\mu)
\end{array}\right)
\equiv
\left(\begin{array}{l}
\Ta_1(x_\mu)\ n^{\al}\\
\Ta_2(x_\mu)\ m^{A}\\
\Ta_3(x_\mu)
\end{array}\right) \,,
\ee
in which the {\it unit} vectors~\footnote{
The two scalar doublet field $\f^\al$ and $\ta^\al$ are parametrised with the same $SO(2)$ rotation angle
$\psi$ since they are both gauged by the ``physical'' $SO(2)$ (Maxwell) field $A_\mu$.}
$n^\al$ and $m^A$ are parametrised as
\be
\label{nal}
n^{\al}=\left(\begin{array}{l}
\cos\psi(x_\mu)\\
\sin\psi(x_\mu)\\
\end{array}\right)\, ,\quad
m^{A}=\left(\begin{array}{l}
\cos\chi(x_\mu)\\
\sin\chi(x_\mu)\\
\end{array}\right)\,.
\ee


The quadratic Skyrme kinetic terms in this parametrisation are
\bea
|D_\mu\f^a|^2&=&|\pa_\mu h|^2+\sin^2h|(A_\mu-\pa_\mu\psi)|^2 \,, \label{Df2}\\
|D_\mu\ta^{\bar a}|^2&=&|\pa_\mu f|^2+\sin^2f\left[|\pa_\mu g|^2+\sin^2g|(A_\mu-\pa_\mu\psi)|^2
+\cos^2g|(B_\mu-\pa_\mu\chi)|^2\right]\label{Dt2} \,,
\eea 
which are both manifestly gauge invariant.

In this parametrisation, the full SCS density \re{SCS} is expressed as
\bea
\label{SCSgaugecomp}
\frac{1}{3!}\Om_{\rm SCS}&=&\vep^{\mu\nu\la}\bigg\{
-2\,\left(\cos f-\frac13\cos^3f
\right)\,(\pa_\la\sin^2g)\,\pa_\mu\psi\,\pa_\nu\chi\nonumber\\
&&\quad\quad\quad+\frac13\,\cos^3f\,(A_\la G_{\mu\nu}+B_\la F_{\mu\nu})\nonumber\\
&&\quad\quad\quad-A_\mu B_\nu\,\cos f\left[\pa_\la(\sin^2f\sin^2g)-\pa_\la(\sin^2f\cos^2g)\right]\nonumber\\
&&\quad\quad\quad 
+2\cos f\left[A_\la\,\pa_\mu(\sin^2f\cos^2g)\ \pa_\nu\chi+B_\la\,\pa_\mu(\sin^2f\sin^2g)\ \pa_\nu\psi\right]\bigg\}\,,
\eea
the first line of which is $\om$ in \re{SCS}, which obviously vanishes in the static limit.

Like the CS density, the SCS density \re{SCSgaugecomp} is not manifestly gauge invariant --
but the equations of motion resulting from both the CS and SCS densities are gauge invariant. This is becsuse the SCS
density results $via$ the one-step descent of the total divergence density $\vr^{(D)}$, \re{vr2}, which is also gauge invariant.
The argument is identical to the case of the CS density which results from the Chern-Pontryagin (CP) density which like  $\vr^{(D)}$
is both gauge-invariant and total-divergence.

In this parametrisation, the full Lagrangian can be expressed as
\bea
\label{full}
{\cal L}=&-&\frac14\,|F_{\mu\nu}|^2-\frac14\,|G_{\mu\nu}|^2\nonumber\\
&+&\frac12\eta^2\left\{|\pa_\mu h|^2+\sin^2h|(A_\mu-\pa_\mu\psi)|^2\right\}\nonumber\\
&+&\frac12\eta^2\left\{|\pa_\mu f|^2+\sin^2f\left[|\pa_\mu g|^2+\sin^2g|(A_\mu-\pa_\mu\psi)|^2
+\cos^2g|(B_\mu-\pa_\mu\chi)|^2\right]\right\}\nonumber\\
&+&\eta\ka\,\Om_{\rm SCS} - \eta^4\, V[\f^a]   -\eta^4\,V[\ta^{\bar a}]  \,,
\eea
in which $\Om_{\rm SCS}$ is given by \re{SCSgaugecomp}.

\subsection{Contractions of the SCS density: Ensuing contracted models}

\subsubsection{Contraction of the auxiliary $O(5)$ scalar to $O(3)$}

Our criterion is to carry out such contractions of
the auxiliary $O(5)$ Skyrme scalar $\ta^{\bar a}$, so that the remaining degrees of freedom are those of an
$O(3)$ Skyrme scalar like the scalar $\f^a$ parametrising the Abelian gauged Skyrmion. 

The necessity for this arises from the fact that the kinetic term $|D_\mu\ta^{\bar a}|^2$ of the
auxiliary $O(5)$ Skyrme scalar in \re{Lta} cannot sustain finite energy {\it topologically} stable solutions on
$\R^2$, since on $\R^2$, only an $O(3)$ Skyrme scalar can support such solutions.

The two natural such contractions are
\bea
\ta^{\bar a}=(\ta^\al,\ta^A,\ta^5)\stackrel{\ta^\al=0}\longrightarrow(\ta^A,\ta^5)&\Leftrightarrow&
g(x_\mu)=0 \,, \label{conf1}\\
\ta^{\bar a}=(\ta^\al,\ta^A,\ta^5)\stackrel{\ta^A=0}\longrightarrow(\ta^\al,\ta^5)&\Leftrightarrow&
g(x_\mu)=\frac{\pi}{2}\,.\label{conf2}
\eea

The resulting contracted SCS densities are
\bea
\Om_{\rm SCS}^{(I)}&=&2\,\vep^{\mu\nu\la}\bigg\{
\cos^3f\,(A_\la G_{\mu\nu}+B_\la F_{\mu\nu})-2\,(\pa_\la\cos^3f)A_\mu(B_\nu-2\pa_\nu\chi)\bigg\} \,, \label{S1}\\
\Om_{\rm SCS}^{(II)}&=&2\,\vep^{\mu\nu\la}\bigg\{
\cos^3f\,(A_\la G_{\mu\nu}+B_\la F_{\mu\nu})-2\,(\pa_\la\cos^3f)B_\mu(A_\nu-2\pa_\nu\psi)\bigg\}\,,\label{S2}
\eea
respectively.

Likewise, the contractions \re{conf1}-\re{conf2} lead to the truncated versions the kinetic term \re{Dt2}
\bea
|D_\mu\ta^{\bar a}|^2&=&|\pa_\mu f|^2+\sin^2f|\,(B_\mu-\pa_\mu\chi)|^2 \,, \label{Dt1x}\\
|D_\mu\ta^{\bar a}|^2&=&|\pa_\mu f|^2+\sin^2f\,|(A_\mu-\pa_\mu\psi)|^2 \,, \label{Dt2x}
\eea
respectively.

While these two contractions are equivalent as far as the SCS density is concerned,
there is however a clear disbalance between them. As seen from \re{S1} and \re{S2}, the roles of the
soliton gauge field $A_\mu$ and the role of the auxiliary gauge field $B_\mu$ are reversed.
But due to the appearance of $A_\mu$ and the absence of $B_\mu$
in the Lagrangian \re{Af}, this symmetry between $A_\mu$ and $B_\mu$
is lost in the full system. The effect of this disbalance will appear in the calculation of global charges
and in the numerical constructions.


The two truncated Lagrangians in constraint compliant parametrisation, pertaining to the contractions
\re{S1} and \re{S2} of the SCS density, can be obtained by setting $g(x_\mu)=0$ and $g(x_\mu)=\pi/2$ in
\re{full} and replacing $\Om_{\rm SCS}$ there by $\Om_{\rm SCS}^{(I)}$ and $\Om_{\rm SCS}^{(II)}$
in turn.



\subsubsection{Further contraction of $g(x_\mu)=0$ model: The $A_\mu=B_\mu$ model}
Having proposed the contracted $g(x_\mu=0)$ model above featuring the gauge fields
$(A_\mu,B_\mu)$, the $O(3)$ Skyrme scalar $\f^a=(\f^\al,\f^3)$ and the auxiliary $O(3)$ scalar $(\ta^A,\ta^5)$,
we now carry out a further contraction aimed at identifying $(\ta^A,\ta^5)$ with $(\f^\al,\f^3)$, $i.e.$,
reducing to a model with only one $O(3)$ scalar. This amounts to setting
\[
\f^1=\ta^3, \ \f^2=\ta^4,\ \f^3=\ta^5\quad\Rightarrow\quad h(x_\mu)=f(x_\mu)\quad {\rm and}\quad 
\psi(x_\mu)=\chi(x_\mu) \,,
\]
whence the Lagrangian \re{full} reduces to
\bea
\label{full11}
{\cal L}^{(I)}=&-&\frac14\,|F_{\mu\nu}|^2-\frac14\,|G_{\mu\nu}|^2\nonumber\\
&+&\frac12\eta^2\left\{2|\pa_\mu f|^2+\sin^2f[|(A_\mu-\pa_\mu\chi)|^2+|(B_\mu-\pa_\mu\chi)|^2]\right\}
\nonumber\\&+&\eta\ka\,\Om_{\rm SCS}^{(I)} - 2\eta^4 V[f] \,,
\eea
which is problematic from the point of view of gauge invariance since the two distinct gauge fields $A_\mu$
and $B_\mu$ transform as
\be
A_\mu\to A_\mu+\pa_\mu\chi\quad{\rm and}\quad B_\mu\to B_\mu+\pa_\mu\chi \,,
\ee
$i.e.$, two distinct Abelian gauge fields transforming with the same gauge rotation function $\chi(x_\mu)$.

This problem can be resolved by making the identification
\be
B_\mu=A_\mu\,,
\ee
resulting in the truncated model
\bea
\label{full111}
{\cal L}^{(0)}=&-&\frac14\,|F_{\mu\nu}|^2
+\frac12 \eta^2\left\{|\pa_\mu f|^2+\sin^2f
|(A_\mu-\pa_\mu\chi)|^2\right\}\nonumber\\
&+&4\eta\ka\,\vep^{\la\mu\nu}\,A_\la(F_{\mu\nu}\,\cos^3f-2\,\pa_\nu\chi\,\pa_\mu\cos^3f) - \eta^4 V[f] \,.
\eea
    
The marked feature of this model is that it is described by one Abelian (Maxwell) field $A_\mu$, and
a single $O(3)$ Skyrme scalar $\f^a$, like $e.g.,$ in the
Callan-Witten~\cite{Callan:1983nx} model which features the Maxwell field and a single $O(4)$ Skyrme scalar.

\section{The Ansatz, equations of motion and global charges}

\subsection{Imposition of static azimuthal symmetry}
Imposition of azimuthal symmetry on the $O(3)$ Skyrme scalar $\f^a=(\f^\al,\f^3)$ is expressed
simply by restricting the functions $\F_1(x_\mu)$ and $\F_2(x_\mu)$ appearing in \re{ccf} as
\be
\label{axF}
\F_1(x_\mu)=\sin h(r)\,,\quad\F_2(x_\mu)=\cos h(r) \,.
\ee
Imposition of azimuthal symmetry on the
$O(5)$ Skyrme scalar $\ta^{\bar a}=(\ta^\al,\ta^A,\ta^5)\ ,\ \al=1,2;\ A=3,4$ is expressed
by restricting the functions $\Ta_1(x_\mu), \Ta_2(x_\mu)$ and $\Ta_3(x_\mu)$ appearing in \re{cct} as
\be
\label{axTa}
\Ta_1(x_\mu)=\sin f(r)\sin g(r)\,,\quad\Ta_2(x_\mu)=\sin f(r)\cos g(r)\,,\quad\Ta_3(x_\mu)=\cos f(r) \,,
\ee
and the functions $\psi(x_\mu)$ and $\chi(x_\mu)$ appearing in the doublet valued functions $n^\al$ and $m^A$ respectively
 in \re{nal} as
\be
\label{psir}
\psi(x_\mu)=n\,\vf\quad{\rm and}\quad\chi(x_\mu)=m\,\vf \,,
\ee
$n$ and $m$ being integers -- the winding numbers  pertaining to the azimuthal angle $\vf$, respectively.

With the usual choice of boundary conditions for sigma models, the corresponding one-dimensional conditions read
\bea
\lim_{r \rightarrow 0} h(r) =\pi \,, \qquad \lim_{r \rightarrow\infty} h(r) &=&0 \,,\label{bvh1}\\
\lim_{r \rightarrow 0} f(r) =\pi \, ,\qquad \lim_{r \rightarrow\infty} f(r) &=&0 \,, \label{bvf1}
\eea
consistent with the finiteness of the energy, as well as with regularity at the origin. This choice of boundary values
means that we are excluding the choice of symmetry-breaking
potentials.

Imposition of azimuthal symmetry on the Abelian gauge fields $A_\mu=(A_i,A_0)$ and $B_\mu=(B_i,B_0)\ \mu=i,0$, is achieved by
\bea
&&A_i=\left(\frac{a(r)-n}{r}\right)(\vep\hat x)_i\quad\Rightarrow\quad F_{ij}=-\frac{a'}{r}\vep_{ij}\, , \label{MaxaxA}
\\
&&A_0=c(r)\qquad\qquad\ \ \Rightarrow\quad\, F_{i0}=c'\,\hat x_i \, , \label{Maxax0A}
\eea
and
\bea
&&B_i=\left(\frac{b(r)-m}{r}\right)(\vep\hat x)_i\quad\Rightarrow\quad G_{ij}=-\frac{b'}{r}\vep_{ij}\, , \label{MaxaxB}\\
&&B_0=d(r)\qquad\qquad\ \ \Rightarrow\quad\, G_{i0}=d'\,\hat x_i \, , \label{Maxax0B}
\eea

\noindent with $\hat x_i \equiv x_i/r$, $i=1,2$ and $(\vep\hat x)_1=\hat x_2 \ , \ (\vep\hat x)_2=-\hat x_1$.

Concerning the boundary values of the magnetic functions $a(r)$ and $b(r)$ parametrising respectively
$A_i$ in \re{MaxaxA} and  $B_i$ in \re{MaxaxB},
consistent with regularity at the origin and finite energy, are
\bea
 \lim_{r \rightarrow 0} a(r) =n \,, \qquad   \lim_{r \rightarrow\infty} a(r) &=&a_\infty \,,\label{bva}\\
 \lim_{r \rightarrow 0} b(r) =m \,, \qquad   \lim_{r \rightarrow\infty} b(r) &=&b_\infty \,, \label{bvb}
\eea
while the corresponding boundary values of the electric functions $c(r)$ and $d(r)$, parametrising respectively
$A_0$ in \re{Maxax0A} and $B_0$ in \re{Maxax0B} are
\bea
\lim_{r \rightarrow 0} c'(r) =0 \,, \qquad \lim_{r \rightarrow\infty} c(r) &=&c_\infty \,,\label{bvc}\\
\lim_{r \rightarrow 0} d'(r) =0 \, ,\qquad \lim_{r \rightarrow\infty} d(r) &=&d_\infty \,, \label{bvd}
\eea
$a_\infty$ and $b_\infty$  can, in general, be considered as free parameters (taking values in certain ranges). Following our previous results in~\cite{Navarro-Lerida:2016omj,Navarro-Lerida:2018giv,Navarro-Lerida:2018siw}, it can be seen that the asymptotic values of $c(r)$ and $d(r)$, denoted by $c_\infty$ and $d_\infty$, respectively, are (numerically) determined once $a_\infty$ and $b_\infty$ are set.

\subsection{Equations of motion and global charges}

Apart from the total mass-energy $E$ and the angular momentum $J$,
the other global charges are the gauge
deformed ``baryon number'' $q$
together with 
the electric charge(s) $Q_e$ and $Q_g$ 
pertaining to the Maxwell field $A_\mu$
and the auxiliary field $B_\mu$, respectively. 
Having opted for the ``pion mass'' potential \re{pot}, this model cannot describe any 
``baryon number'' change. So our considerations are restricted to the global {\it electric charge(s)} and global
{\it angular momentum}.

The definitions of
electric charges involve the SCS term, while the definition of the angular momentum is independent of the SCS term but
is strongly influenced by the electric charges and hence also by the SCS term.
 The energy density $T_{00}$ is
also independent of the SCS term but the solutions are influenced by it.
The total mass-energy of solutions is defined as usual
\begin{eqnarray}
E=\int T_{00} d^2x \,.
\end{eqnarray}
The angular momentum $J$ is defined   in terms of the angular momentum density
\be
{\cal J}\stackrel{\rm def.}=-(\vep x)_i\,T_{0i}=T_{0\vf} \,, \label{dJ}
\ee
by the integral
\be
\label{J}
J=\int\,{\cal J}\,d^2x=2\pi\,\int_0^\infty \,{\cal J}rdr \,.
\ee

The definition(s) of the electric charge(s) hinge on the gauge field, Maxwell $A_\mu$ and auxiliary field $B_\mu$,
equations. 
One starts with
the Maxwell equations
\bea
&&\pa_\mu\,F^{\mu\tau}+2\cdot3!\eta\ka\,\vep^{\tau\mu\nu}\,
\cos f\left[\frac13\,\cos^2 f\,G_{\mu\nu}-\pa_\mu(\sin^2f\cos^2g)\,(B_\nu-\pa_\nu\chi)\right]
=j^\tau \,, \label{Max}\\
&&j^\tau=-\eta^2\left[\sin^2h+\sin^2f\sin^2g\right](A^\tau-\pa^\tau\psi)\, ,
\label{jt}
\eea
%
together
with  the auxiliary gauge field equations of the system \re{L},
namely the variational equation w.r.t. $B_\tau$,
\bea
&&\pa_\mu\,G^{\mu\tau}+2\cdot3!\eta\ka\,\vep^{\tau\mu\nu}\,
\cos f\left[\frac13\,\cos^2 f\,F_{\mu\nu}-\pa_\mu(\sin^2f\sin^2g)\,(A_\nu-\pa_\nu\psi)\right]
=\tilde j^\tau \,, \label{aux}\\
&&\tilde j^\tau=-\eta^2\,\sin^2f\cos^2g(B^\tau-\pa^\tau\chi)\, ,
\label{tjt}
\eea
which are as expected gauge invariant.

In the definition of the electric charges,
it is important to note that the vector valued quantities with strength $\ka$
in \re{Max} and in \re{aux} are not divergenceless, as a result
$j^\tau$ and $\tilde j^\tau$ in \re{Max}-\re{jt} and \re{aux}-\re{tjt} are not divergenceless, so
their temporal components $j_0$ and $\tilde j_0$ cannot qualify as {\it conserved charge} densities.

To define conserved ``electric'' charges, one takes recourse to the Noether Theorem, by defining the respective
Noether currents
\bea
j_{\rm N}^\tau&=&\frac{\pa{\cal L}}{\pa(\pa_\tau\psi)} \,,
\label{jN}\\
\tilde j_{\rm N}^\tau&=&\frac{\pa{\cal L}}{\pa(\pa_\tau\chi)} \,,
\label{tjN}
\eea
corresponding to the two Abelian gauge parameters $\psi$ and $\chi$, respectively.

After a straightforward calculation, one has
\be
\label{jNA}
j_{\rm N}^\tau=j^\tau+2\cdot3!\eta\ka\vep^{\tau\mu\nu}\left[\left(\cos f-\frac13\,\cos^3f\right)(\pa_\mu\sin^2g)\pa_\nu\chi
-\cos f\pa_\mu(\sin^2f\,\sin^2g)B_\nu\right] \,,
\ee
pertaining to the Maxwell field, and
\be
\label{jNB}
\tilde j_{\rm N}^\tau=\tilde j^\tau-2\cdot3!\eta\ka\vep^{\tau\mu\nu}\left[\left(\cos f-\frac13\,\cos^3f\right)(\pa_\mu\sin^2g)\pa_\nu\psi
+\cos f\pa_\mu(\sin^2f\,\cos^2g)A_\nu\right] \,,
\ee
pertaining to the auxiliary gauge field.

Substituting $j^\tau$ from \re{Max} into $j_{\rm N}^\tau$ given by \re{jNA}, one finds the ``improved'' Maxwell equation
\be
\label{impMax}
\pa_\mu F^{\mu\tau}+2^3\eta\ka\,\vep^{\tau\mu\nu}\,\pa_\mu\left[\cos^3f(B_\nu-\cos^2g\,\pa_\nu\chi)\right]=j_{\rm N}^\tau \,,
\ee
the L.H.S. of which is divergenceless, such that the temporal component of the current $j_{\rm N}^\tau$ can define the
conserved (Maxwell) electric charge
\be
\label{QeA}
Q_e=\frac{1}{2\pi}\int\,j_{\rm N}^0\,d^2x \,.
\ee

Likewise substituting $\tilde j^\tau$ from \re{aux} into $\tilde j_{\rm N}^\tau$ given by \re{jNB}, one finds the ``improved'' auxiliary gauge field equation
\be
\label{impaux}
\pa_\mu G^{\mu\tau}+2^3\eta\ka\,\vep^{\tau\mu\nu}\,\pa_\mu\left[\cos^3f(A_\nu-\sin^2g\,\pa_\nu\psi)\right]=\tilde j_{\rm N}^\tau \,,
\ee
the L.H.S. of which is divergenceless, 
such that the temporal component of the current $\tilde j_{\rm N}^\tau$ can define the
conserved (auxiliary) electric charge
\be
\label{QeB}
Q_g=\frac{1}{2\pi}\int\,\tilde j_{\rm N}^0\,d^2x \,.
\ee

Note that the functions $f, g, \psi$ and $\chi$ appearing in the expressions given in this
Subsection, {\bf 3.2}, are valid before imposition of the symmetry stated in Section {\bf 3.1} above.

For the explicit evaluation of the electric charges \re{QeA} and \re{QeB}, we impose symmetry on the temporal
components of the Noether currents \re{impMax} and \re{impaux} yielding
\bea
j_{\rm N}^0&=&-\frac1r\frac{d}{dr}(rc')-{8\eta\ka}\frac1r\frac{d}{dr}[\cos^3f(b-m\sin^2g)]\,, \label{jN0}\\
\tilde j_{\rm N}^0&=&-\frac1r\frac{d}{dr}(rd')-{8\eta\ka}\frac1r\frac{d}{dr}[\cos^3f(a-n\cos^2g)]\,.\label{tjN0}
\eea

Substituting \re{jN0} and \re{tjN0} in the integrals \re{QeA} and \re{QeB} respectively, we have
\bea
Q_e&=&-8\eta\ka (b_\infty-m \sin^2g_\infty +m \cos^2g_0) \,,
\label{Qe}
\\
Q_g&=&-8\eta\ka (a_\infty -n \cos^2 g_\infty + n \sin^2 g_0)\,,
\label{Qg}
\eea
which can be evaluated if we know the boundary conditions of $g(r)$ (where $g_0=\displaystyle \lim_{r\to 0} g(r)$ and $g_\infty=\displaystyle \lim_{r\to \infty} g(r)$). But since we have not found any
  solutions with non-constant $g(r)$, it is not possible to evaluate $Q_e$ and $Q_g$ in the general case.

After imposition of symmetry, the angular momentum density ${\cal J}$ reduces to
the one dimensional quantity 
\be
\label{calJ}
{\cal J}=(a'c'+b'd')+\eta^2\,a\,c[\sin^2h+\sin^2f\sin^2g]+\eta^2\,d\,b\sin^2f\cos^2g \,,
\ee
depending on the radial coordinate $r$.
One remarks that, 
while $\cal J$ does not have a contribution from the (anomalous) SCS term, the integral \re{J} does so by virtue of the
equations of motion. 

Subjecting the Maxwell equation \re{Max}-\re{jt} and the auxiliary field equation \re{aux}-\re{tjt} to symmetry
we have, respectively
\bea
\frac1r\frac{d}{dr}\left\{(r\,c')+12\,\eta\ka\cos f\left[\frac23\,b'\,\cos^2f-b(\sin^2f\cos^2g)'\right]\right\}
&=&\eta^2\,c\,(\sin^2h+\sin^2f\sin^2g) \,, \quad \quad \quad   \label{Maxx}\\
\frac1r\frac{d}{dr}\left\{(r\,d')+12\,\eta\ka\cos f\left[\frac23\,a'\,\cos^2f-a(\sin^2f\sin^2g)'\right]\right\}
&=&\eta^2\,d\,\sin^2f\cos^2g\, ,    \label{auxx}
\eea
leading to 
\be
\label{calJx}
{\cal J}=\frac1r\frac{d}{dr}\left[r(ac'+bd')+8\eta\ka(a\,b\,\cos^3f)\right]\,,
\ee
which remarkably has no dependence on the Skyrme function $h(r)$.

The integral \re{J} can be directly evaluated to yield
\be
\label{Jx}
J=16\pi \eta \ka (a_\infty\,b_\infty+n\,m)\, .
\ee



\section{Specific models and the results}

The various models considered in this work do not
appear to possess 
 closed-form solutions, and thus we relied on numerical methods to solve the field equations.
The field equations result in second-order ordinary differential equations
(ODEs) for the functions 
$a$, $b$,
$c$, $d$, 
$f$, and $h$, 
which
are solved subject to a set of boundary conditions compatible with finiteness of the energy and
regularity of the solutions, as stated in Section {\bf 3.1}. 

For all reported solutions,  the systems of six  coupled ODEs were solved by using the software
package COLSYS
which employs a collocation
method for boundary-value ordinary differential equations and a damped Newton method
of quasi-linearization  \cite{COLSYS}.
Also, in the numerics we set $\eta=1$ by using a rescaling of the radial coordinate,
such that the only input parameters are  the constant 
$\kappa$
multiplying the SCS term in (\ref{Lta}), the constant 
$\mu$
multiplying the potentials in (\ref{pots})
and the winding numbers $n$ and $m$
in the gauge field Ansatz (\ref{MaxaxA}) and (\ref{MaxaxB}), together with the asymptotic values of the magnetic gauge potentials $a_\infty$ and $b_\infty$.

 
\subsection{
The contracted $g(x_\mu)=0$ model $I$ and $g(x_\mu)=\pi/2$  model $II$}

\begin{figure}[ht!]
\begin{center}
 \includegraphics[height=.34\textwidth, angle =0 ]{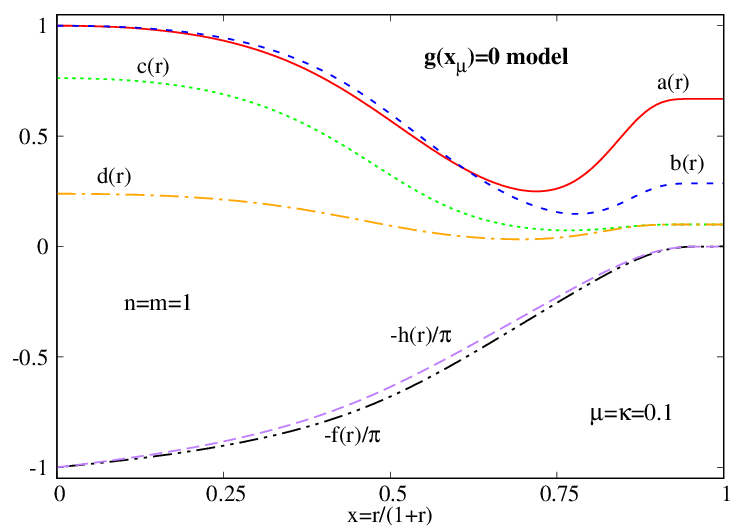} 
 \includegraphics[height=.34\textwidth, angle =0 ]{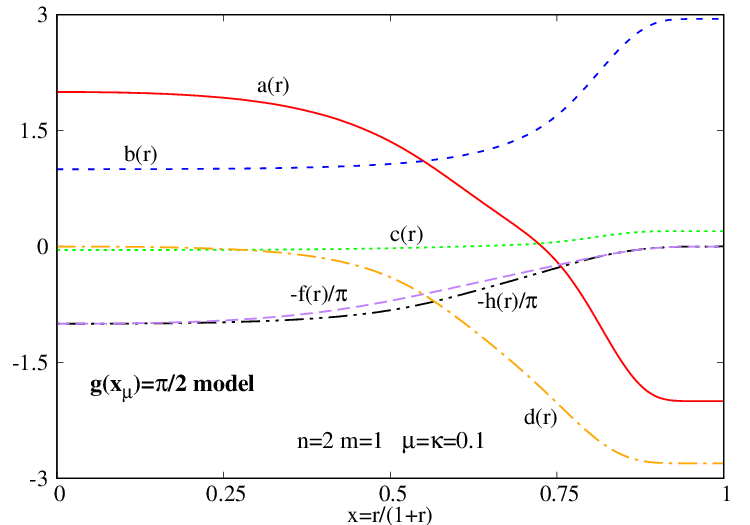}
\end{center}
\caption{  
The profiles of a typical solution are shown for the  contracted $g(x_\mu)=0$
(left panel) and 
 $g(x_\mu)=\pi/2$ (right panel)
models.
}
\label{profile}
\end{figure}

The case 
$g(x_\mu)=0$
corresponds to a vanishing  $\Ta_1$-function in the 
O(5) Skyrme Ansatz (\ref{axTa}), 
while the functions $\Ta_2$ and $\Ta_3$ are nontrivial.
The profile of a typical solution is shown in Fig. \ref{profile} (left panel).
As one can see, the scalar functions 
$f(r)$
and $h(r)$
together with the gauge potentials
$a(r)$,
$c(r)$
and
$b(r)$,
$d(r)$
interpolate smoothly between the
(fixed) values at the origin
and some asymptotic values at infinity.
  Note that, as discussed in Section {\bf 3.1},
the values at infinity of the magnetic gauge potentials, $a_\infty$ and $b_\infty$,
are not fixed a priori but are free parameters, the asymptotic values of $c(r)$ and $d(r)$, $c_\infty$ and $d_\infty$, resulting from the numerical output. 

The functions show a behaviour at $r=0$ given by
\begin{eqnarray}
&&
a(r)=n + a_2 r^2 + O(r^4) \,  ,~~~~~~~~  c(r)= c_0 + O(r^2) \, ,
\label{exp_or_g0_ac}
\\
&&
b(r)= m +b_2 r^2 + O(r^4) \,  ,~~~~~~~~  d(r)=d_0 + O(r^2) \, ,
\label{exp_or_g0_bd}
\\ 
&&
f(r) =\pi +f_m r^m + O(r^{m+2}) \, ,~~~  h(r) =\pi +h_n r^n + O(r^{n+2}) \, ,
\label{exp_or_g0_fh}
\end{eqnarray}
(with $a_2,b_2,c_0,d_0,f_m,h_n$ constant parameters),
 and an asymptotic behaviour given by
\bea
a(r)=a_\infty + a_1 r^{-1/2} e^{-8\kappa r} + \dots  \, ,  &
c(r) = c_\infty +  b_1 r^{-3/2} e^{-8\kappa r} + \dots  \, ,
\label{exp_inf_g0_ac}
\\b(r)=b_\infty + b_1 r^{-1/2} e^{-8\kappa r} + \dots  \, ,  &
d(r) = d_\infty +  a_1 r^{-3/2} e^{-8\kappa r} + \dots  \, ,
\label{exp_inf_g0_bd}
\\f(r)=f_1 r^{-1/2} e^{-\sqrt{\mu -d_\infty^2} \, r} + \dots  \, ,  &
h(r) = h_1 r^{-1/2} e^{-\sqrt{\mu -c_\infty^2} \, r} + \dots  \, ,
\label{exp_inf_g0_fh}
\eea
for $\eta=1$.

The expressions of the electric charges $Q_e$ and $Q_g$ result by setting $g=0$
in the 
general expressions
(\ref{Qe}), 
 (\ref{Qg})
above, with
\bea
Q_e&=& \int\,j_N^0\,r\,dr=
-8\eta\ka(b_\infty+m) \, ,
\label{Qetrunc}
\\
Q_g&=& \int\,\tilde j_N^0\,r\,dr=
-8\eta\ka(a_\infty-n)
\label{Qgtrunc} \, ,
\eea
while
the angular momentum
$J$
is given by (\ref{Jx}).

\begin{figure}[ht!]
\begin{center}
 \includegraphics[height=.34\textwidth, angle =0 ]{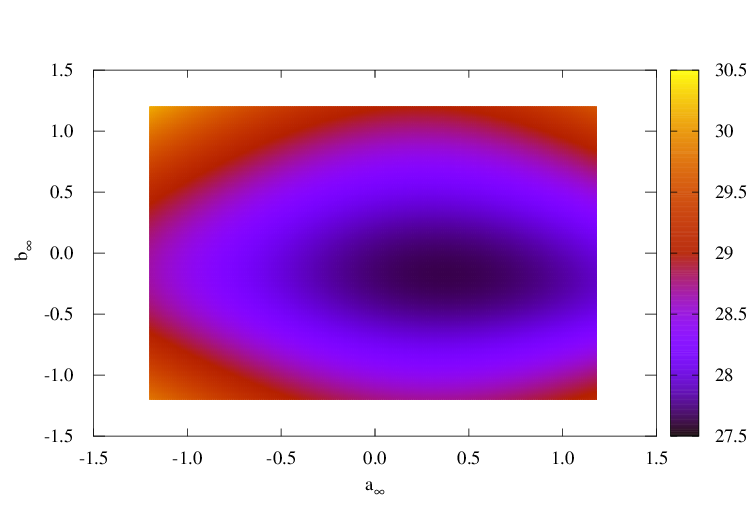}
\end{center}
\caption{  
The mass-energy $E$ is shown as a function of the asymptotic values of the 
magnetic gauge potentials $(a_\infty,b_\infty)$ for solutions
with $n=m=1$, $\mu=0.1$, and $\kappa=-0.1$. 
Similar results have been found for other values of $\kappa$.
}
\label{g0-sup}
\end{figure}
 
\begin{figure}[ht!]
\begin{center}
 \includegraphics[height=.34\textwidth, angle =0 ]{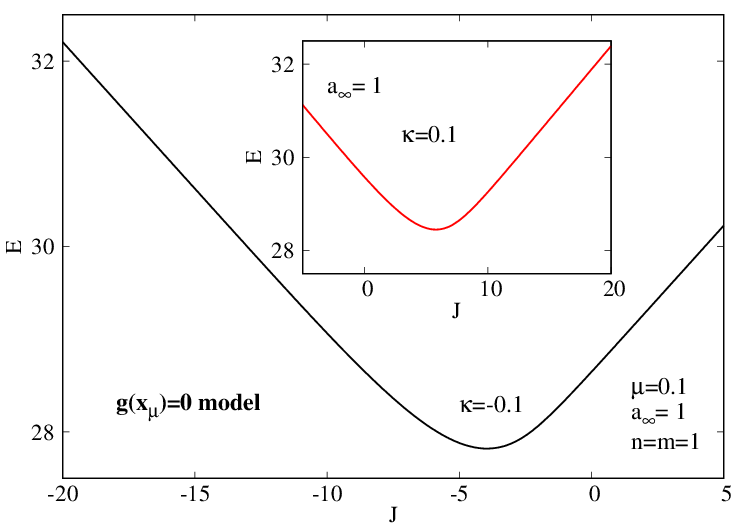}
 \includegraphics[height=.34\textwidth, angle =0 ]{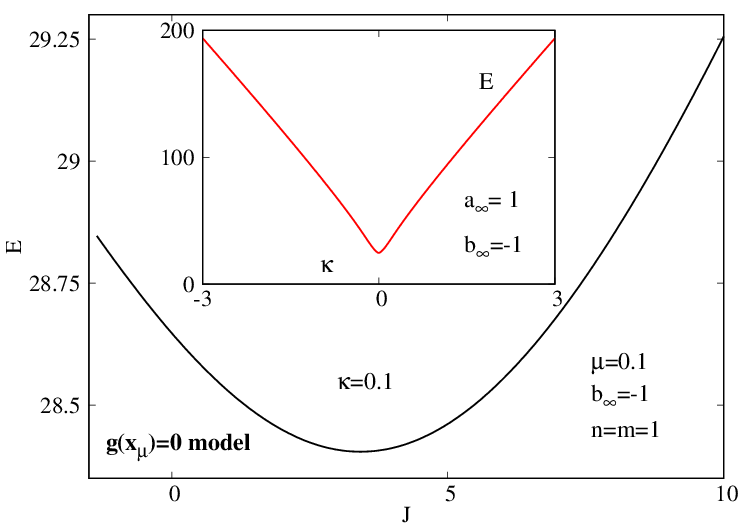}
\end{center}
\caption{  
The $(E,J)$-diagram of the contracted $g(x_\mu)=0$ model
is shown for solutions with   fixed  $\kappa$ and a fixed
magnetostatic potentials $a_\infty$ (left panel)
and  $b_\infty$ (right panel).
The insets show how the energy varies as a function of $J$ with fixed  $a_\infty$ for opposite $\kappa$ (left) and as a function of $\kappa$
for configurations with fixed  $a_\infty$ and $b_\infty$. (In all plots $\mu=0.1$.)
}
\label{g0-fig3}
\end{figure}
\begin{figure}[ht!]
\begin{center}
 \includegraphics[height=.34\textwidth, angle =0 ]{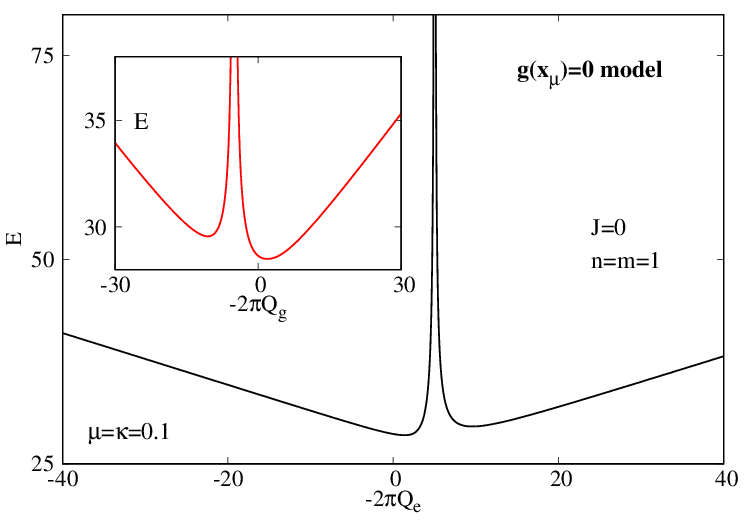}
\end{center}
\caption{  
The  mass-energy $E$ of solutions
is shown as a function of the electric and the Noether charges for 
 solutions for the contracted $g(x_\mu)=0$ model with vanishing total angular momentum.
}
\label{g0-fig4}
\end{figure}

%
The colour map in Fig. \ref{g0-sup}
displays the total energy $E$ as a function of 
$(a_\infty,b_\infty)$ for a set of solutions with fixed $\kappa$ and $\mu$. 
Clearly, the minimum of energy does not occur at 
$Q_e=Q_g=0$ ($i.e.$, at $a_\infty=n$, $b_\infty=-m$). Further insight can be seen in Fig. \ref{g0-fig3}
where we display the  ($E,J$)-diagram
for a set of solutions with  
fixed $a_\infty$ (left panel) and $b_\infty$ (right panel).
One remarks that the  
$E(J)\neq E(-J)$,
 while the minimal value of $E$ is
approached for some non-zero $J$.
Solutions with $J=0$
exist as well, the dependence of mass-energy $E$ on
the electric charges $(Q_e, Q_g)$
being shown in Fig. \ref{g0-fig4}.

\begin{figure}[ht!]
\begin{center}
 \includegraphics[height=.34\textwidth, angle =0 ]{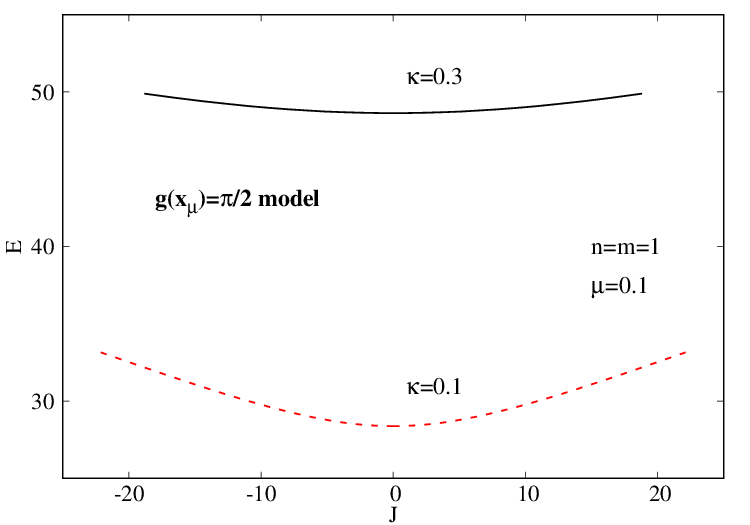}
\end{center}
\caption{  
The $(E,J)$-diagram  of the contracted $g(x_\mu)=\pi/2$ model.
}
\label{gpi2-fig2}
\end{figure}

\medskip
Similar solutions are found when 
taking 
$g(x_\mu)=\pi/2$ 
in (\ref{axTa}),
$i.e.$ for a vanishing
 $\Ta_2$-Skyrme function,
while the functions $\Ta_1$ and $\Ta_3$ are nontrivial.
As with the $g(x_\mu)=0$ case, 
the expressions of 
$Q_e$ and $Q_g$ 
result from the
general relations
(\ref{Qe}), 
 (\ref{Qg})
with
\begin{eqnarray}
\label{Qe111}
Q_e &=&-8\eta\ka(b_\infty-m),
\\
Q_g&=&0,
\end{eqnarray}
and
 $J$
still given by (\ref{Jx}).

A typical solution of the $g(x_\mu)=\pi/2$ model is shown in Fig. \ref{profile} (right panel),
while 
the $(E-J)$ (energy-angular momentum) diagram is displayed
in Fig. \ref{gpi2-fig2} for two fixed values of $\kappa$. 
 While the expansions at the origin coincide with those of the $g(x_\mu)=0$ mode, (\ref{exp_or_g0_ac})-(\ref{exp_or_g0_fh}), the asymptotic behaviour deserves further explanation. The equations for the auxiliary gauge potential 
($i.e.$, equations for functions $b$ and $d$) 
are in this case total derivatives, so they can be integrated one order:
\begin{eqnarray}
\label{eq_b_g_pi2}
\left(\frac{1}{r}b'\right)' + 8 \kappa \eta (c \cos^3 f )' &=&0\, ,
\\
\left(r d'\right)' + 8 \kappa \eta (a \cos^3 f )' &=&0\, . \label{eq_d_g_pi2}
\end{eqnarray}
Then regularity at the origin together with (\ref{bvf1}) and (\ref{bva}) lead to the simple expression
\be
a_\infty + n =0 \, .
\ee
Therefore $a_\infty$ is fixed for this contracted model, 
$b_\infty$ being the only free parameter, 
once the values of $\mu$, $\kappa$, $n$, and $m$ are given. 
Owing to that, the system of field equations can be effectively reduced to a system for $a$, $c$, $f$, and $h$ only. 
The asymptotic behaviour for these functions is (with $\eta=1$)
\bea
&&a(r)=-n + a_1 r^{1/2} e^{-8\kappa r} + {\hat a}_1 r^{-1} e^{-2\sqrt{\mu -c_\infty^2}\, r}    \dots  \, , \label{exp_inf_gpi2_a}
\\ 
&&c(r)=c_\infty + c_1 r^{1/2} e^{-8\kappa r} + {\hat c}_1 r^{-1} e^{-2\sqrt{\mu -c_\infty^2}\, r}    \dots  \, , \label{exp_inf_gpi2_c}
\\
&&f(r)=f_1 r^{-1/2} e^{-\sqrt{\mu -c_\infty^2} \, r} + \dots  \, , \ \ \ 
h(r) = h_1 r^{-1/2} e^{-\sqrt{\mu -c_\infty^2} \, r} + \dots  \, .
\label{exp_inf_gpi2_fh}
\eea
 
As seen in Fig.  \ref{gpi2-fig2} 
 the $g(x_\mu)=\pi/2$ model has $E(J)=E(-J)$,
$i.e.$
 the asymmetry in the $(E,J)$-plot found in the $g(x_\mu)=0$ model is lost.

\subsection{The contracted $A_\mu=B_\mu$ model}

This model is described by the Lagrangian
(\ref{full111})
and it is interesting in itself. 
 The corresponding Maxwell equations are
\bea
&&\pa_\mu F^{\mu\tau}+8\eta\ka\,\vep^{\tau\mu\nu}[F_{\mu\nu}\cos^3f+\pa_\mu\cos^3f(A_\nu-\pa_\nu\chi)]=j^\tau \, ,
\label{MaxAB}\\
&&
j^\tau=-2\eta^2\,\sin^2f(A^\tau-\pa^\tau\chi)\,.\nonumber
\eea

Defining 
\be
\label{NoetherAB}
j_N^\tau\stackrel{\rm def.}=\frac{\pa{\cal L}}{\pa(\pa_\tau\chi)}=j^\tau+8\eta\ka\,\vep^{\tau\mu\nu}
A_\nu\pa_\mu\cos^3f \, ,
\ee
\re{MaxAB} is expressed as
\be
\pa_\mu F^{\mu\tau}+8\eta\ka\,\vep^{\tau\mu\nu}\pa_\mu[(2A_\nu-\pa_\nu\chi)\cos^3f]=j_N^\tau \, , \label{MaxABimptd}
\ee
the L.H.S. of which is a total divergence, and hence $j_N^0$ can be identified as the electric charge density
\bea
j_N^0&=&-\pa_i\,F_{i0}+8\eta\ka\,\vep_{ij}\,
\left[\cos^3 f\,F_{ij}+(\pa_i\cos^3f)\,(2A_j-\pa_j\chi)\right] \, ,
\label{Maxximp0}
\eea
which subject to the Ans\"atze in Section {\bf 4}, 
 the current \re{Maxximp0} reduces to
\be
j_N^0=-\frac1r\frac{d}{dr}\left\{(rc')+8\eta\ka[(2a-n)\cos^3f]\right\} \, , \label{GLAB}
\ee
leading to the electric charge 
\be
Q_e=\int_0^\infty\,j_N^0\,rdr=-16\eta\ka\,a_\infty \, . \label{QeAB}
\ee

The angular momentum density is
\be
\label{calJAB}
{\cal J}=a'\,c'+\eta^2\,a\,c\,\sin^2f\,.
\ee
Imposing symmetry on the Maxwell equation \re{MaxAB} we get
\be
\label{redMaxAB}
\frac1r\frac{d}{dr}(rc')+\frac{8\eta\kappa}{r}[2\,a'\,\cos^3f+a(\cos^3f)']=\eta^2\,c\,\sin^2f\,,
\ee
and finally substituting \re{redMaxAB} into \re{calJAB} we have 
\be
\label{calJABfin}
{\cal J}=\frac1r\frac{d}{dr}[rac'+8\eta \kappa a^2 \cos^3f]\,,
\ee
so that the angular momentum is
\be
\label{JABfin}
J=2\pi\int_0^\infty {\cal J}\,r\,dr=2\pi\bigg[rac'+8\eta \kappa a^2 \cos^3f\bigg]_{r=o}^{r=\infty}=16\pi\eta\kappa (a_\infty^2+n^2) \,.
\ee

\begin{figure}[ht!]
\begin{center}
 \includegraphics[height=.34\textwidth, angle =0 ]{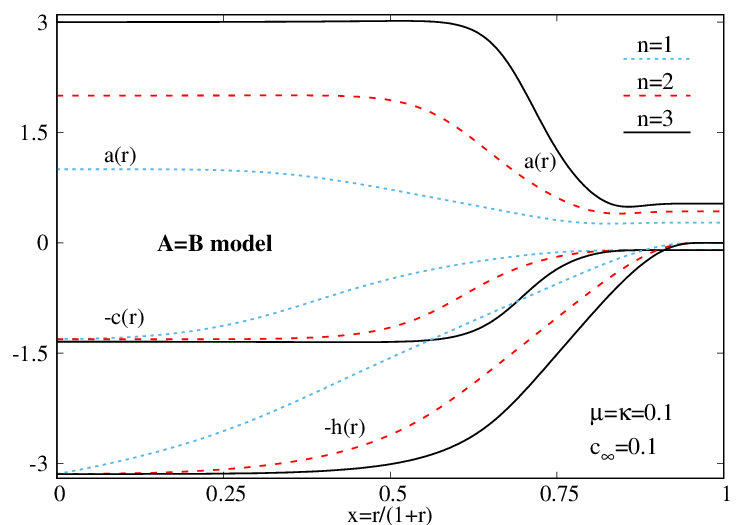}
\end{center}
\caption{ 
The profile of  typical solutions of the  $A_\mu=B_\mu$ model with three values of
the vorticity number $n$.
}
\label{AB-fig1}
\end{figure}

\begin{figure}[ht!]
\begin{center}
 \includegraphics[height=.34\textwidth, angle =0 ]{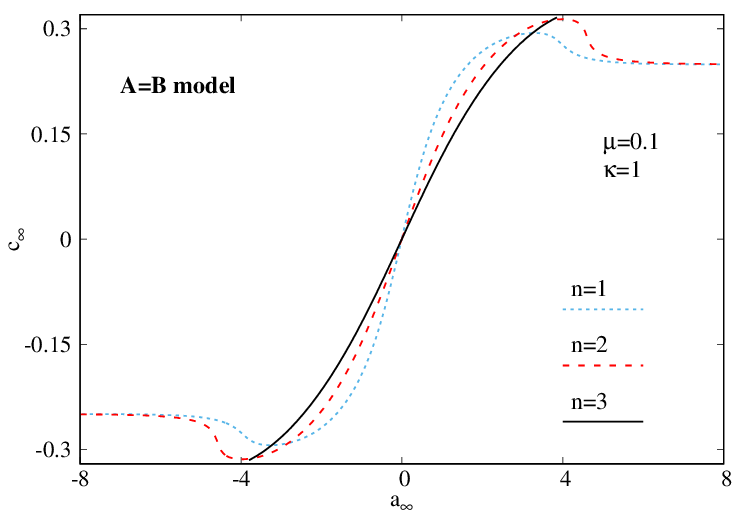}
 \includegraphics[height=.34\textwidth, angle =0 ]{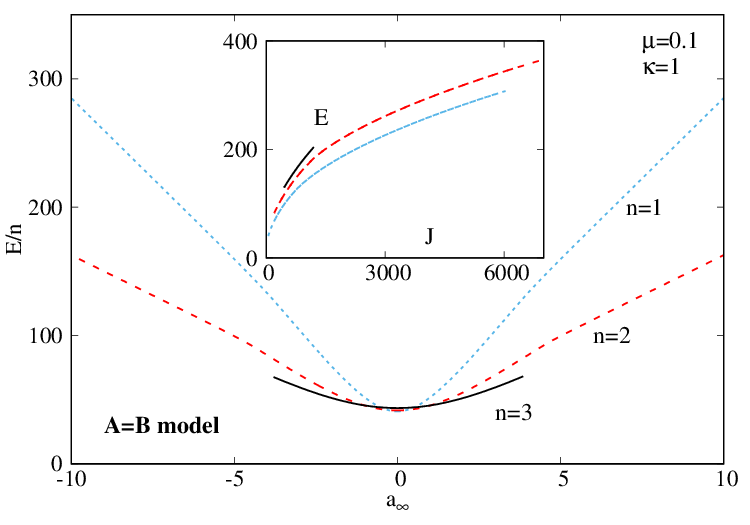}
\end{center}
\caption{   
The electrostatic potential at infinity $c_\infty$
(left panel)
and the mass-energy 
(right panel)
are shown as a function of the 
magnetostatic potential at infinity $a_\infty$ for solutions of the $(A_\mu=B_\mu)$-model
with several values of the winding number $n$.
The inset shows the $(E,J)$-diagram for the same solutions.
}
\label{AB-fig2}
\end{figure}

The profile of a typical solution of the $A_\mu=B_\mu$
model is shown in Fig. \ref{AB-fig1}.
 The functions interpolate between their behavior at the origin given by
 \bea
a(r)=n + a_2 r^2 + O(r^4) \, ,  &c(r)= c_0 + O(r^2) \, , &f(r) =\pi +f_n r^n + O(r^{n+2})  \, ,
\label{exp_or_A=B_acf}
\eea
and their asymptotic behaviour
\bea
a(r)=a_\infty + a_1 r^{1/2} e^{-16\kappa r} + \dots  \, ,  &
c(r) = c_\infty +  a_1 r^{-1/2} e^{-16\kappa r} + \dots  \, , \label{exp_inf_A=B_ac}
\\f(r)=f_1 r^{-1/2} e^{-\sqrt{\mu -c_\infty^2} \, r} + \dots  \, ,
\label{exp_inf_A=B_f}
\eea
for $\eta=1$. 
Again, only $a_\infty$ is a free parameter, $c_\infty$ being determined once $a_\infty$ is given (or viceversa), 
while $f_1$ is also fixed by numerics.

In  Fig. \ref{AB-fig2}   several quantities of interest
 are displayed
as a function of the
asymptotic value of the magnetic
 potential at infinity $a_\infty$. 
The results there were found for three values of the winding number $n$ and a fixed value of the coupling constant
$\kappa$ and of the constant of the potential $\mu$.
 From \re{JABfin} it is clear that the angular momentum $J$ has a definite sign (that of $\kappa$). Also from the inset in Fig.~\ref{AB-fig2}, we can see that for this model the slope of the $E(J)$ is strictly positive, not showing the behaviour of the $g(x_\mu)=0$ model. In addition, solutions with $J=0$ are not allowed for non-vanishing $\kappa$.

\section{Summary and discussion}

We have studied the $SO(2)$ gauged $O(3)$ Skyrme model in $2+1$ dimensions,
in the presence of a SCS term, and in the absence of the
(usual) CS term. The SCS density is
defined in terms of $SO(2)\times SO(2)$ gauge fields, one of these being
the ``physical'' Maxwell field $A_\mu$  gauging the $O(3)$ Skyrme scalar $\f^a\ , a=1,2,3,$
supporting the Skyrmion (soliton) in the absence of the SCS term, and the other being the auxiliary
Abelian field $B_\mu$. In addition, the definition of the SCS term involves the auxiliary $O(5)$ scalar
 $\ta^{\bar a}\ ,\bar a= 1,2,3,4,5$.

This is the first concrete application of a SCS density in a gauged Skyrme model,
where in addition to defining the global
charges pertaining to the two Abelian gauge fields and the angular momentum,
a detailed numerical construction of the      
solutions supporting these quantities is provided.
The $F^2$ Maxwell gauged $O(3)$ Skyrmion in $2+1$ dimensions in the presence of the ($AF$)
CS term has been studied extensively,
in Refs.~\cite{Navarro-Lerida:2016omj,Navarro-Lerida:2018giv,Navarro-Lerida:2018siw}, and the $F^4$
Maxwell gauged $O(5)$ Skyrmion in $4+1$ dimensions in the presence of the ($AFF$)
CS term has been studied  in Ref.~\cite{Navarro-Lerida:2020jft}. In both these studies,
some remarkable properties of the solutions were observed, namely the occurrence of negative slopes in
the energy versus electric charge $(E,Q_e)$ and energy versus angular momentum $(E,J)$ plots.
In addition to these, the value of the ``baryon number'' of the gauged Skyrmion departed 
from the {\it integer valued winding number}.

The aim here is to reveal to what extent the properties resulting from the CS dynamics are present also for the
case of SCS dynamics. Should the dynamical effects of the SCS be qualitatively similar to those of the
usual CS, this would indicate that in even spacetime dimensions, where a CS term is not defined, the appropriate
SCS term can be employed to result in similar effects, namely the occurence of negative slopes in $(E,Q)$
and $(E,J)$ curves, and baryon number decay. In the present $2+1$ dimensional ``test model'', we have
indeed observed the presence of negative slopes in the $(E,Q)$ and $(E,J)$ curves, though the decay of ``baryon
number'' turns out to be absent here. Inspite of the absence of baryon number decay here in
$2+1$ dimensions, we expect that this property will be present in higher dimensions. Indeed, this obstacle
is restricted to $2+1$ dimensions where the Bogomol'nyi type lower bound requires the presence of a potential
like \re{pot}, while in higher dimensions, $e.g.,$ in $4+1$ dimensions concerned with Abelian Chern-Simons
gauging of the O(6) Skyrmion in Ref.~\cite{Navarro-Lerida:2020jft}, 
the Bogomol'nyi type lower bounds do not require the presence of a potential like \re{AppC} which would have
caused the vanishing of the integral of the second term in \re{vr2x}. Indeed in that case the ``pion mass''
potential $V\simeq(1-\f^6)$ was used in the numerical construction.




We have studied two main contractions
of the $O(5)$ SCS density parametrised by two distinct Abelian gauge
fields $A_\mu$ and $B_\mu$ and the auxiliary $O(5)$ Skyrme scalar $\ta^{\bar a}\ ;\ \bar a=(\al,A,5)\ ; \al=1,2\ ;A=3,4$,
down to two effective $O(3)$ Skyrme models. Here, $A_\mu$ is the Maxwell field gauging the Skyrmion prior to introducing the
SCS term, and $B_\mu$ is an auxiliary Abelian field sustaining the SCS term. In the first of these contractions
$(i)$ one sets $\ta^\al=0$ resulting in the effective O(3) auxiliary Skyrme scalar $\ta^a=(\ta^A,\ta^5)$, and in the other $(ii)$ $\ta^A=0$
resulting in $\ta^a=(\ta^\al,\ta^5)$~\footnote{Contractions of the auxiliary Skyrme scalar to
$\ta^a=(\ta^\al,0,0)$ gauged with an Abelian field, and contractions to $\ta^a=(\ta^I,0,0)\ ,\ I=1,2,3$ gauged with $SO(3)$
are eschewed, since the dynamical effects of these on the Skyrmion are trivial~\cite{Tchrakian:2021xzy}.}.
This means that in the case $(i)$ the SCS term is parametrised by the auxiliary
Abelian field $B_\mu$, while in the case $(ii)$ it is parametrised by the Maxwell field $A_\mu$ itself. We have observed that
the SCS term resulting from the first contraction $(i)$ replicates the effects of the CS dynamics, while the SCS
term pertaining to the second contraction $(ii)$ does not. Does this mean that the role of the auxiliary Abelian field cannot
be performed by the Maxwell field?

To answer this question we recall a third contraction, $(iii)$, studied above, namely when  the contracted auxiliary scalar of case $(i)$  
$\ta^a=(\ta^A,\ta^5)$, is identified with $\f^a=(\f^\al,\f^3)\equiv(\ta^A,\ta^5)$
that results in the identification $A_\mu=B_\mu$. It turns out that also in this case the properties of CS dynamics are
not replicated, from which it might be concluded that the role of a distinct auxiliary gauge field $B_\mu$ cannot be 
taken by the Maxwell field $A_\mu$. This is in stark contrast to the situation in $3+1$ dimensions in the
context of the Callan-Witten~\cite{Callan:1983nx} (CW) anomaly, which is conceptually similar to a SCS density
for an $SO(2)$ gauged Skyrmion in $3+1$ dimensions. (Indeed, the definition of that SCS term
is inspired by the CW anomaly.) Such a study was carried out recently
in Ref.~\cite{Navarro-Lerida:2023hbv}, where it was found that the effect of the CW term on the
$U(1)$ gauged Skyrmion is similar to the effect of the SCS term employed in the $2+1$ dimensional model studied here, namely it results in the negative slopes in $(E,Q_e)$ and $(E,J)$ plots.
But the CW term in $3+1$ dimensions is parametrised by the Maxwell field and the $O(4)$ Skyrme scalar $\f^a\ ,\ a=1,2,3,4$ (or the group element $U$) as the ones that parametrise the Lagrangian of the Abelian gauged Skyrmion. In other words, the roles of the auxiliary gauge field and Skyrme scalar are the same as the ones parametrising the Skyrmion prior to introducing the CW term to the Lagrangian.
That there is such a contrast between odd and even dimensional spacetimes is perhaps
not surprising.

 Our results indicate that the effects observed due to the presence of the CS term in the Lagrangian
supporting the gauge Skyrmion in $2+1$ dimensions, are qualitatively observed also when the CS term is replaced by a
SCS density. Hence, we expect that employing a SCS term in the Lagrangian of a gauged Skyrmion
in even dimensions will lead to effects qualitatively similar to those observed with CS dynamics in odd dimensions. Most importantly, this concerns $3+1$ dimensions which is under active consideration.

In addition to the role of prototype played by the $2+1$ dimensonal model here, it may be
appropriate to mention that theories involving a CS term featuring two distinct Abelian
fields, also occur in some applications in condensed matter physics, in gauged Skyrme models, in $e.g.,$
Ref.~\cite{Palumbo:2014lqa}, and in gauged Higgs models, in $e.g.,$ Ref.~\cite{DeLima:2022qka}.

\section*{Acknowledgements}
F.N.-L. gratefully acknowledges support  from MICINN under project PID2021-125617NB-I00 ``QuasiMode" and also from Santander-UCM under project PR44/21‐29910.
The work of E.R.
is supported  by the  Center for Research and Development in Mathematics and Applications (CIDMA) through the Portuguese Foundation for Science and Technology (FCT -- Fundac\~ao para a Ci\^encia e a Tecnologia), references  UIDB/04106/2020 and UIDP/04106/2020.
E.R. also acknowledges support  from the projects 
CERN/FIS-PAR/0027/2019,
PTDC/FIS-AST/3041/2020,  
CERN/FIS-PAR/0024/2021 
and 
2022.04560.PTDC.  
This work has further been supported by  the  European  Union's  Horizon  2020  research  and  innovation  (RISE) programme H2020-MSCA-RISE-2017 Grant No.~FunFiCO-777740 and by the European Horizon Europe staff exchange (SE) programme HORIZON-MSCA-2021-SE-01 Grant No.~NewFunFiCO-101086251.

\appendix
\section{The Skyrme--Chern-Simons density}
\setcounter{equation}{0}
\renewcommand{\theequation}{A.\arabic{equation}}

In this Appendix, we give a brief sketch of the construction of the SCS term employed here. The starting point is
$\vr^{(4)}$, the $D=4$ analogue of \re{vr2} given $e.g.,$ in \cite{Tchrakian:2015pka,Tchrakian:2021xzy}, 
pertaining to the $SO(4)$ gauged $O(5)$ model in $D=4$, namely
\bea
\vr^{(4)}
&=&\vr_0^{(4)}+\pa_l\Om_l^{(4)}\label{vr2y}
\eea
where $\Om_l^{(4)}$ in \re{vr2y} is given by

\bea
\label{Oml4}
\Om_l^{(4)}&=&\ta^5\vep_{ijkl}\vep^{{f}{g}{a}{b}}\ta^5\bigg\{A_k^{{f}{g}}\,\ta^{{a}}
\left(\pa_j\ta^{{b}}-\frac12A_j\ta^{{b}}\right)
+\frac12\,F_{ij}^{{f}{g}}\ta^{{a}}D_k\ta^{{b}}
\nonumber\\
&&\qquad\qquad 
+\frac14\left(1-\frac13(\ta^5)^2\right)A_k^{f g}\left[\pa_iA_j^{a b}
+\frac23(A_iA_j)^{a b}\right]\bigg\}
\eea
in which $\ta^{a}\  ;\ a=1,2,3,4$ together with $\ta^5$ defines the $O(5)$ Skyrme scalar 
$\ta^{\bar a}\ ;\ \bar a=(a,\ta^5)\ ,\ $
referred to as the auxiliary Skyrme scalar in Section {\bf 2}, which is subject to the constraint
\[
|\ta^{\bar{a}}|^2=|\ta^{{a}}|^2+(\ta^5)^2=1\,.
\]

The quantity $\Om$ in the definition \re{SCS} of the SCS density is the $l=4$ component of $\Om_l^{(4)}$ in
\re{Oml4}. The density $\Om_{l=4}^{(4)}$, which is now defined in three ``spacetime'' dimensions, pertains to the
$SO(4)$ gauge connection. 

However, the SCS term employed in Section {\bf 2} is that resulting from the
contraction~\footnote{The other contractions of $SO(4)$
to $SO(2)$ or $SO(3)$ lead to SCS densities that depend only on one power of the connection 
and in the static limit depend only on the time component $A_0$. This makes the CS like effects trivial.}
of the $SO(4)$ connection
$A_\mu^{{a}{b}}\  ;\ a=(\al,A)\ ;\ \al=1,2\ ;\quad A=3,4$\ ; to the  $SO(2)\times SO(2)$ connections
\be
\label{contr}
A_\mu^{\al\bt}=A_\mu\,\vep^{\al\bt} \,,\quad A_\mu^{AB}=B_\mu\,\vep^{AB} \ ,\quad A_\mu^{\al A}=0\ ,
\ee
in terms of the two Abelian connections $A_\mu$ and $B_\mu$.

After the contraction \re{contr}, the $l=4$ component of $\Om_l^{(4)}$ in \re{Oml4} results in $\Om$ displayed
in \re{SCSAB}, which when added to $\om$ is the full SCS density considered in this work.

Finally, in the constraint compliant parametrisation given in Section {\bf 2.1} the full SCS density is
given by \re{SCSgaugecomp}.

In the defninition of the model in Section {\bf 2}, $A_\mu$ is taken to be the Maxwell field which gauges
the $O(3)$ Skyrme scalar $\f^a$ that supports the soliton, and $B_\mu$ is referred to as the auxiliary gauge
field.  Thus, the two Skyrme scalars $\f^a$ and $\ta^{\bar{a}}$ see each other through the Maxwell field $A_\mu$.

\begin{small}
\end{small}
\end{document}